\documentclass[aps,prl,twocolumn,showpacs,preprintnumbers,amsmath,amssymb,10pt]{revtex4-1}

\usepackage{graphicx}
\usepackage{dcolumn}
\usepackage{bm}
\usepackage{hyperref}
\usepackage{epstopdf}
\usepackage{amsmath}
\usepackage{url}      
\usepackage{textcomp}
\usepackage{color}
\usepackage{booktabs} 
\graphicspath{{./}{Fig/}}

\newcommand{\om}{\omega} 
 
\newcommand{\ep}{\epsilon} 
\renewcommand{\Re}{{\it Re} } 
\renewcommand{\Im}{{\it Im} } 
\newcommand{\beq}{\begin{equation}}
\newcommand{\eeq}{\end{equation}}

\begin{document}
\author{Honghua U. Yang}
\author{Markus B. Raschke}
\email{markus.raschke@colorado.edu}
\affiliation{Department of Physics, Department of Chemistry, and JILA, University of Colorado, Boulder, CO 80309, USA}%

\title{Optical gradient force nano-imaging and -spectroscopy}

\date{\today}

\begin{abstract}
Nanoscale forces play an important role in different scanning probe microscopies, most notably atomic force microscopy (AFM). 
In contrast, in scanning near-field optical microscopy (SNOM) a light-induced coupled local optical polarization between tip and sample is typically detected by scattering to the far field.
Measurements of the optical gradient force associated with that optical near-field excitation would offer a novel optical scanning probe modality. 
Here we provide a generalized theory of optical gradient force nano-imaging and -spectroscopy.
We quantify magnitude and distance dependence of the optical gradient force and its spectral response.
We show that the optical gradient force is dispersive for single particle electronic and vibrational resonances, distinct from recent claims of its experimental observation.
In contrast, the force can be absorptive for collective resonances.
We provide a guidance for its measurements and distinction from competing processes such as thermal expansion. 

\end{abstract}

\maketitle
\begin{figure}[b]
\begin{center}
    \includegraphics[width=\columnwidth]{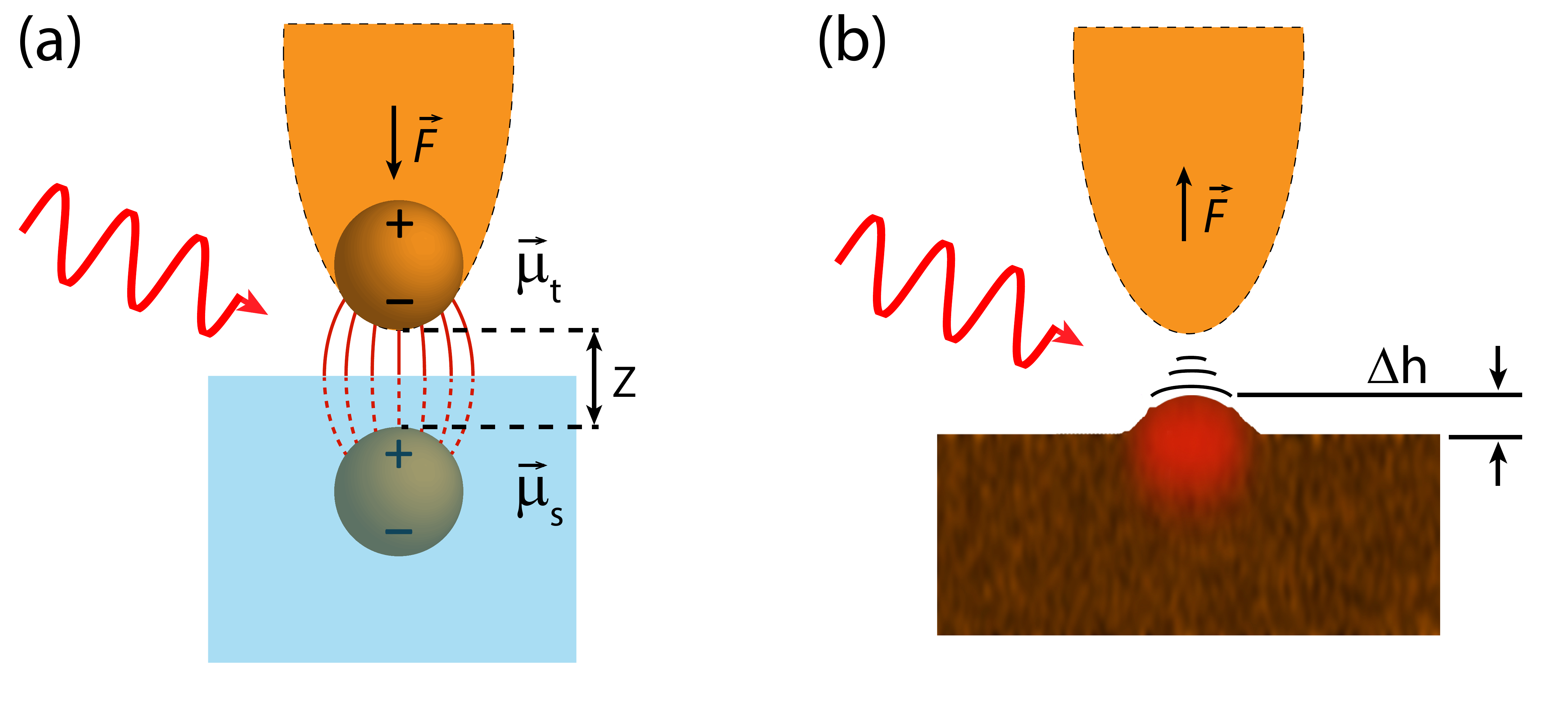}
    \caption[Mechanisms]{
(a) Optical gradient force induced by excitation of coupled optical polarization between tip ($\mu_t$) and sample ($\mu_s$), with resulting force $F \propto \mu_t \mu_s$.
(b) The optical excitation is accompanied by absorption, with the resulting thermal expansion $\Delta h$ gives rise to a competing force reaction of the AFM tip. 
}
\label{Mechanisms}
\end{center}
\end{figure}

Control and enhancement of light-matter interaction via radiation pressure and optical gradient forces form the basis of cavity optomechanics and optical tweezers \cite{Anetsberger2009, Berthelot2014}.
Fundamentally related, the tip-sample interaction in electrostatic, magnetic, or atomic force microscopy is based on classical electromagnetic or quantum electrodynamic effects (Van der Waals/Casimir interaction) \cite{Giessibl2003, Emmrich2015}.
Under light illumination, the induced spatially localized coupled optical polarization between the tip and the sample forms the basis of scanning near-field optical microscopy (SNOM)  \cite{Betzig1991, Zenhausern1995}, with the near-field signal typically detected by far-field scattering. 
However, the coupled optical polarization is expected to also give rise to an optical gradient force between the tip and the sample as illustrated in Fig.~\ref{Mechanisms}a \cite{Depasse1992, Zhu1997, Kohlgraf-Owens2014}. 
Its mechanical detection by means of atomic force microscopy (AFM) techniques has been proposed as an promising alternative way for scanning probe optical nano-imaging and -spectroscopy \cite{Satoh2010,  Rajapaksa2010a, Rajapaksa2011a, DeAngelis2012, Kohlgraf-Owens2014}.

Here, we provide a generalized theoretical description of the optical gradient force between a scanning tip and sample as an imaging contrast mechanism.
The spectroscopic response of the optical gradient force is found to be dispersive for molecular electronic or vibrational and other single particle excitations, in contrast to recent experimental claims of gradient force nano-spectroscopy \cite{Rajapaksa2010a, Jahng2014}.
We find that only collective polariton resonances can give rise to absorptive spectral force profile.
The effect is distinct from the accompanying thermal expansion due to optical absorption (Fig.~\ref{Mechanisms}b), which results in absorptive resonance spectra in all cases \cite{Dazzi2005, Dazzi2012, Lu2014}.
While the optical gradient force is near the thermal cantilever noise limit in a room temperature AFM, the effect should be detectable under improved force detection or cryogenic conditions as established in cavity opto-mechanics \cite{Anetsberger2009}.

Recent experimental efforts on measuring the optical near-field gradient force have not been conclusive.
For example, a force image of a metal bowtie antenna on resonance at $\lambda = 1550$ nm has been interpreted to result from an optical gradient force \cite{Kohoutek2011}.
In contrast, similar force images on gold split ring resonators were attributed to thermal expansion \cite{Katzenmeyer2014}.
For molecular electronic resonance spectroscopy an absorptive response has been observed and attributed to the optical gradient force \cite{Rajapaksa2010a, Jahng2014}, in marked contrast to the predictions in this work.
In contrast, a nearly instantaneous (few ps) response from ultrafast experiments \cite{Jahng2015} and polarization dependence in imaging plasmonic particles \cite{Huang2015} show characteristics of the expected optical gradient force behavior. 
However, both experiments lack spectroscopic information which would be desirable for a unique assignment.
The situation is equally confusing regarding theory. Most studies so far have treated both tip and sample as point dipoles \cite{Rajapaksa2010a, Jahng2014}, even at close proximity where the dipole approximation fails, or only calculated the distance dependence of the force without considering its spectral response \cite{DeAngelis2012}, leaving key open questions on experimental feasibility and distinguishing spectral characteristics.

In this work, we numerically calculate the optical gradient force between the tip and sample from a surface integral of the Maxwell stress tensor $T_{ij}$ using finite element electromagnetic simulation (COMSOL Multiphysics).
The force on the tip is given by
\beq
F_i = \sum_j \int_S T_{ij} \hat{n}_j ds
\eeq
with tip surface $S$, surface normal $\hat{n}$, and 
\beq
T_{ij} = \epsilon_0 \left( E_i E_j  - \frac{1}{2}|E|^2 \delta_{ij} \right) + \frac{1}{\mu_0} \left( B_i B_j -\frac{1}{2}|B|^2 \delta_{ij} \right).
\eeq
Since the radius of tip apex $r \simeq 10 $ nm is much smaller than the laser wavelength $\lambda$, the near-field tip and sample interaction under plane wave illumination can be analyzed in the quasistatic approximation by assuming the probe to be a polarizable sphere with radius $r$.
The resulting field distribution on the sample can be effectively reduced to an image sphere of radius $r$, as shown in Fig.~\ref{Mechanisms}a.
While the exact geometry of tip and sample affects both the details of magnitude and spectral response of the optical gradient force, especially for strong polaritonic resonances, the limiting case of two finite spheres provides enough general insight into the spectral variation of the force spectrum and its distance dependence. 

Considering incident polarization along the tip axis, and rotational symmetry along the $z$-axis allows simulation in reduced dimensions to decrease the computational complexity.
For the field simulation,  a dense mesh with element size of 0.2 nm or less is applied near the spheres, and the distant surrounding is meshed with maximum element size of 20 nm.
A uniform electric field $E = 10^6$ V/m along the $z$ direction is applied.
This electric field corresponds to an average laser intensity of $I_0 = 1.33 \ {\rm mW}/\mu{\rm m}^2$ as used under typical experiment conditions \cite{Rajapaksa2010a, Berweger2013b}.

We consider two types of resonant processes of the sample: i) single particle molecular electronic and vibrational resonances, and ii) collective plasmonic resonances.
The results are readily generalizable to samples with other types of resonances including phonons, excitons, and related polaritons.
For the case of an electronic resonance, a sample consisting of Rhodamine 6G (R6G) dye molecules is modeled as a single harmonic oscillator for its dielectric function $\epsilon(\omega) = \epsilon_1(\omega) + i \epsilon_2(\omega)$, with transition energy of 2.42 eV, and line width of 0.41 eV, as a best fit to experimental results. 
A tungsten tip with $r = 10$ nm is used to ensure a flat spectral response of the tip in the relevant visible spectral region. 
For the molecular vibrational resonance, a gold tip sphere is used to interact with a sample sphere (both with $r=10$ nm) of poly(methyle methacrylate) (PMMA) with characteristic carbonyl resonance (C=O) at 1729 cm$^{-1}$.
Finally, to study the case of plasmonic resonance, two silver spheres of $r=10$ nm are modeled as the tip and sample system.

\begin{figure}[tb]
\begin{center}
    \includegraphics[width=\columnwidth]{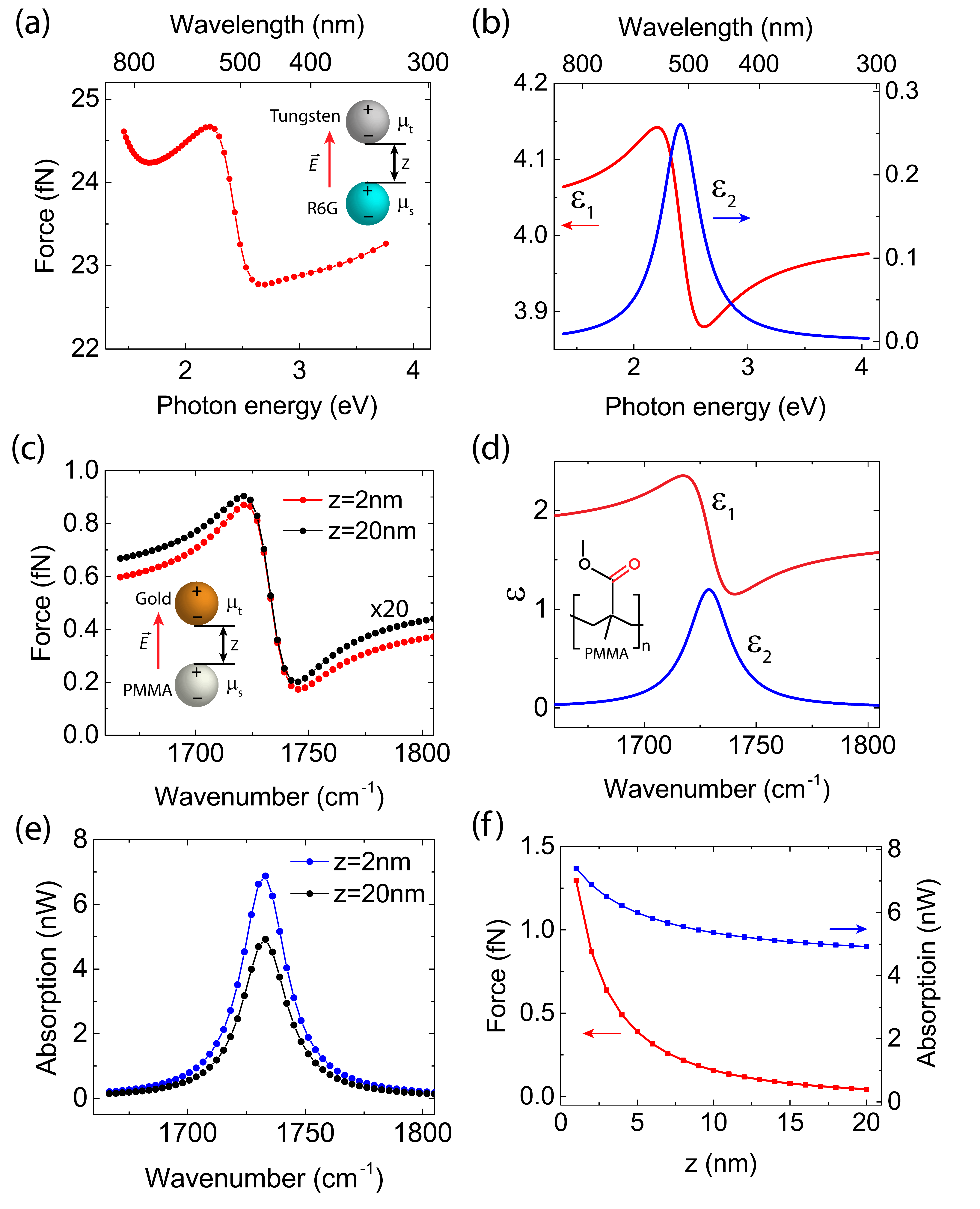}
    \caption[Optical force spectrum of a PMMA and a gold sphere]{
        (a) Electronic resonant force spectrum between a tungsten tip and R6G sample (both of radius $r=10$ nm) subject to an optical field $E = 10^6$ V/m, separated by $z=2$ nm.
        (b) Dielectric function of R6G $\ep(\om) = \ep_1(\om) + i \ep_2(\om)$ modeled as a single harmonic oscillator.
    (c) Corresponding force spectra between a gold sphere and PMMA sample of radii $r=10$ nm.
    Dispersive force spectrum at separation distance $z=20$ nm (multiplied by 20, black) compared to force spectrum at $z=2$ nm (red).
    (d) Dielectric function of PMMA $\ep(\om) = \ep_1(\om) + i \ep_2(\om)$ around the carbonyl resonance at 1729 cm$^{-1}$.
    (e) Simulated PMMA absorption at distances $z=20$ nm (black) and $z=2$ nm (blue).
    (f) The distance dependence of the optical force (red) on resonance shows a complex distance scaling, neither following simple dipole-dipole power law nor exponential scaling.
    The slight increase of optical absorption (blue) is due to local field enhancement in the gap region.
}
\label{PMMA_spec}
\end{center}
\end{figure}
Fig.~\ref{PMMA_spec}a shows the resulting force spectrum between the tungsten tip and R6G sample at a distance of $z=2$ nm.
The force spectrum is dispersive, and follows the trend of the real part of the dielectric function $\epsilon_1(\omega)$ shown in Fig.~\ref{PMMA_spec}b. 
The relatively small force variation between 23 fN to 25 fN across the resonance is due to the large broadband offset as a result of $\ep_1 \gg \ep_2$, which is a characteristic property of the dielectric function for molecular electronic resonances. 

Correspondingly, Fig.~\ref{PMMA_spec}c show the force spectrum of the Au sphere interacting with the carbonyl vibrational resonance of PMMA at separation of $z=20$ nm (black).  
The force increases by $20$ times in magnitude to a peak value of $F \approx 1$ fN when decreasing the distance to $z=2$ nm (red).
Due to the smaller vibrational dipole moment compared to the electronic counter part, the force is much weaker than the case of R6G but exhibits a larger relative spectral variation across the resonance from 0.2 fN to 1 fN (for $z=2$ nm).
Irrespective of distance, and similar to the electronic resonance case, the force spectra are dispersive and follow the real part of the dielectric function of PMMA $\epsilon_1(\om)$ (Fig.~\ref{PMMA_spec}d).
For comparison, Fig.~\ref{PMMA_spec}e shows the absorption spectra at $z=20$ nm and $z=2$ nm in black and blue, respectively. 
In contrast to the force spectra, the absorption spectra show the well known symmetric resonance behavior corresponding to the imaginary part of the dielectric function $\epsilon_2(\om)$.
Except for a change in magnitude, peak position and line shape of both optical gradient force and absorption spectra are invariant with respect to tip-sample distance. 
The distance dependence of the optical force on resonance (red) is shown in Fig.~\ref{PMMA_spec}f.
Due to increasing multipole contributions with decreasing distance, the distance scaling follows neither simple exponential nor power law, and is far more shallow than what one would expect based on a simple dipole-dipole interaction ($\propto z^4$).
The magnitude of absorption (blue) only increases slightly due to local field enhancement with decreasing distance.

\begin{figure}[tb]
    \includegraphics[width=\columnwidth]{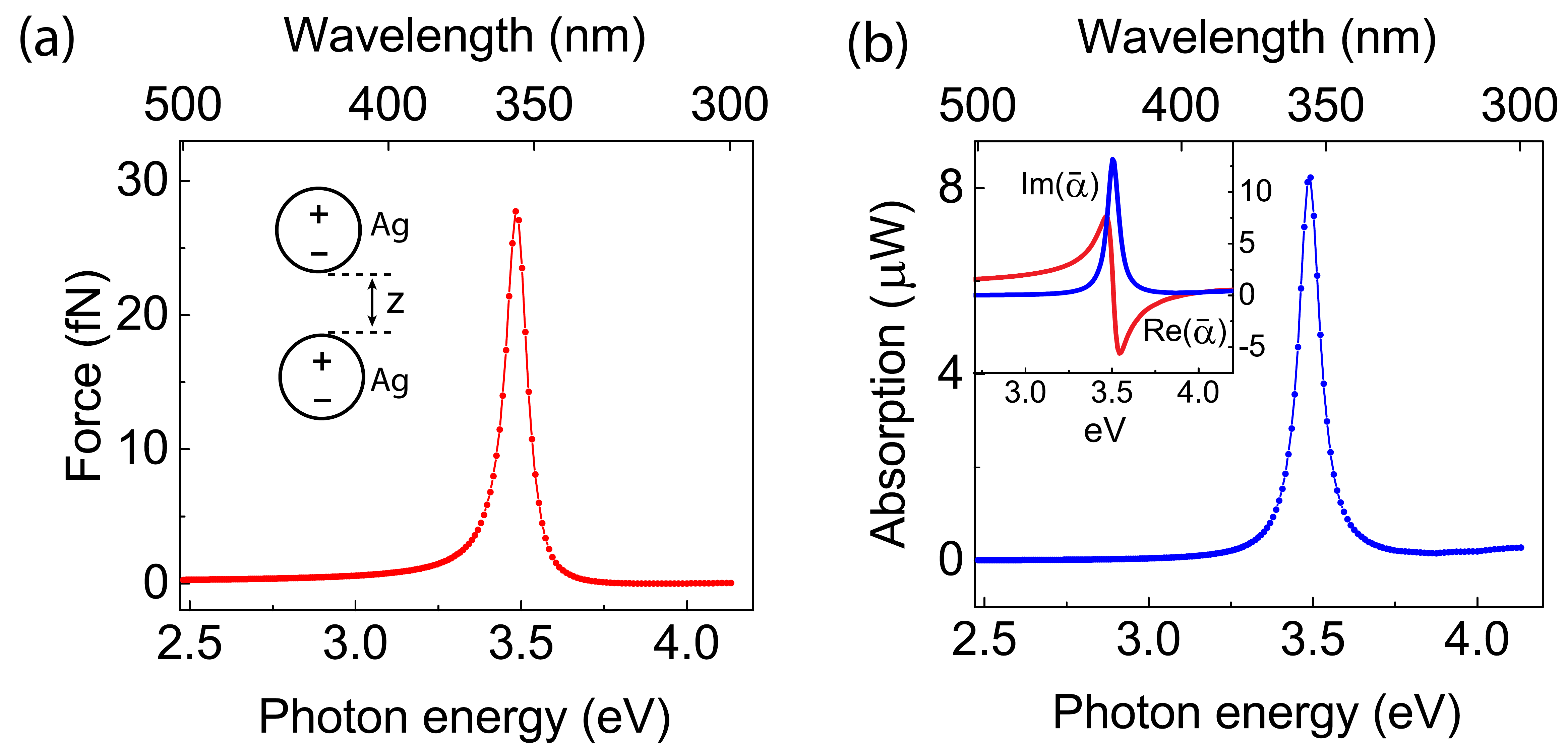}
\caption[Optical force spectrum of two silver particles far apart]{
(a)  Optical gradient force spectrum of two silver spheres with radius $r=10$ nm separated by $z=20$ nm subject to external optical field $E = 10^6$ V/m.
(b) Absorption spectrum of the two spheres with symmetric resonance at 3.5 eV.
Inset: real and imaginary part of the normalized polarizability $\bar{\alpha} = (\ep_{\rm Ag} - 1)/(\ep_{\rm Ag} + 2)$ for a silver sphere.
Force and absorption spectra follow the imaginary part of $\bar{\alpha}(\omega)$.
}
\label{silver_spec}
\end{figure}

Fig.~\ref{silver_spec}a shows the force spectrum of two silver spheres ($r=10$ nm) separated by $z =20$ nm for an applied optical field of $E = 10^6$ V/m. 
The plasmonic resonant force spectrum  with a slight asymmetric peaks at 3.5 eV with $F_{\rm max} \simeq 30$ fN. 
For comparison, Fig.~\ref{silver_spec}b shows the absorption spectrum for the two silver spheres. 
The normalized complex polarizability  of a silver sphere $\bar{\alpha} = (\ep_{\rm Ag} - 1)/(\ep_{\rm Ag}+ 2)$ is shown in the inset.
As can be seen, both optical gradient force and absorption spectra follow the imaginary part of polarizability $\Im(\bar{\alpha})(\omega)$.
\begin{figure*}[tb]
    \includegraphics[width=1.3\columnwidth]{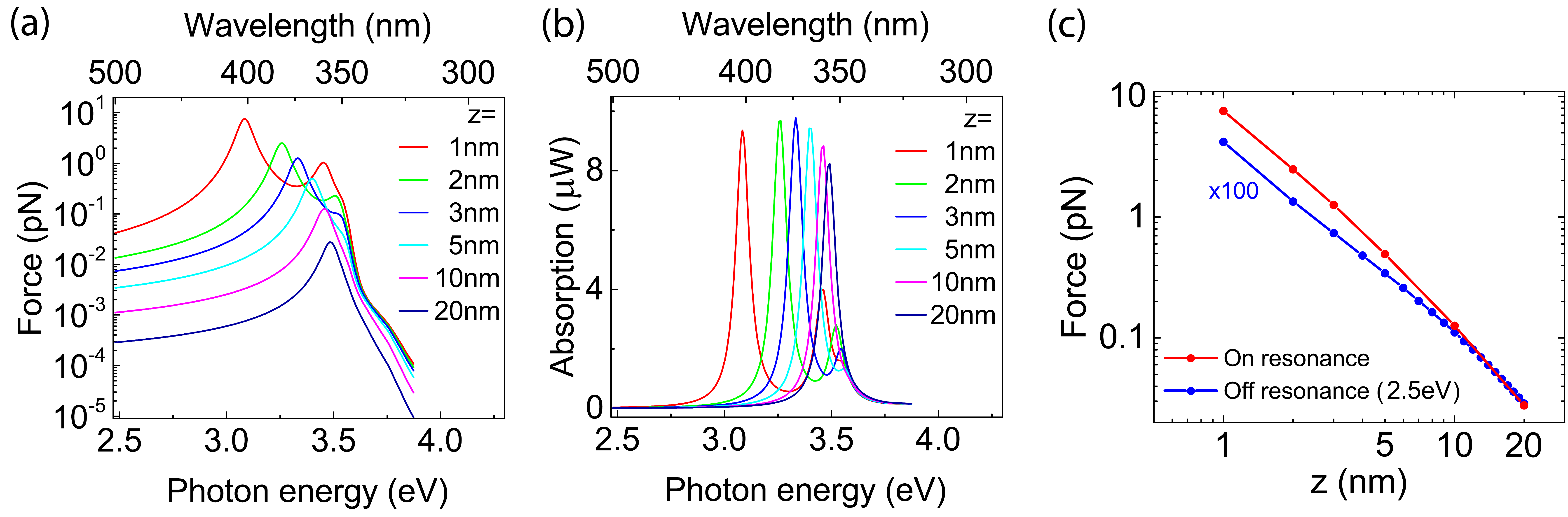}
    \caption[Optical force spectrum of two silver particles separated by different distances]{
        Simulated optical gradient force (a) and absorption (b) spectra for two silver spheres of $r=10$ nm at different separations.
        The peak magnitude of optical force on resonance increases by a factor $\sim 10^3$ from $z=20$ nm to $z=1$ nm. Meanwhile the resonance of the force spectrum red shifts due to plasmonic coupling.
        In comparison, the magnitude of the corresponding absorption spectra has no significant distance dependence.
        (c) The optical gradient force, both on-resonance (red) and off-resonance (blue), follows an power law distance dependence on $z$.
}
\label{dis_dep}
\end{figure*}

Fig.~\ref{dis_dep}a shows the spectral evolution of the force as a function of distance. 
When the separation of the two spheres decreases from $z=20$ nm to $z=1$ nm, the optical gradient force increases by a factor of $\sim 10^3$ reaching $F \approx 10$ pN.
Correspondingly, the resonance frequency of the force spectrum red-shifts due to the plasmonic coupling between the two spheres \cite{Romero2006}.
In contrast, the magnitude of optical absorption does not change significantly with separations as shown in Fig.~\ref{dis_dep}b. 
The distance dependence of the optical gradient force on- (red) and off-resonance (blue) is plotted in Fig.~\ref{dis_dep}c. Both increase with decreasing distance approximately following a power law $\propto z^n$, with $n\approx 2$, i.e., again less steep than the point dipole $z^4$ dependence.

The different spectral behaviors of the optical gradient force can be rationalized based on the dipole-dipole interaction assuming two independently polarized spheres when $z > 2r$, with resulting force proportional to the polarizability,  i.e.,  $F \propto  \Re (\alpha_t \alpha_s^*)$.  
For a non-resonant tip, the force simplifies to $F \propto \Re(\alpha_s)$, describing the dispersive force spectra for a tungsten tip and a R6G sample in the visible, and a gold tip and a PMMA sample in the infrared. 
It is interesting to note that the optical gradient force can in principle also become repulsive for $\Re (\alpha_{s})$ crossing zero.
When tip and sample consist of the same material (e.g., silver), the force spectrum is proportional to the square of the polarization, i.e., $F \propto  \Re (\alpha_t \alpha_s^*) \propto |\alpha_{\rm Ag}|^2$,  and is absorptive when $\Re (\alpha_{\rm Ag})$ crosses zero as for the case for the two plasmonic silver spheres.

As shown above,  the optical gradient force based on a collective plasmonic resonance can reach few pN. 
The force based on a single particle excitation in the form of a molecular resonance is weaker by three orders of magnitude (few fN) under similar tip-sample geometric conditions.
In the following we address the question of AFM sensitivity for its possible detection and distinction from thermal expansion.
The sensitivity of force detection in AFM is limited by thermal fluctuations of the cantilever \cite{Martin1987, Giessibl2003}. 
The smallest detectable force is when the cantilever oscillation amplitude equals that induced by thermal noise given by 
\beq
F_{\rm min} = \left[ \frac{4 k_B T k B}{\om Q} \right]^{1/2},
\eeq
with Boltzmann constant $k_B$, temperature $T$, spring constant $k$, detection bandwidth $B$, and Q factor of the cantilever.
The minimal detectable force gradient due to a change in sample-tip interaction operating in non-contact mode can also be derived as \cite{Martin1987, Giessibl2003}
\beq
F'_{\rm min} = \frac{1}{A_0} \sqrt{ \frac{27  k_B T k B}{\om Q}},
\eeq
with cantilever oscillation amplitude $A_0$.
Using typical values from force detection experiments \cite{Rajapaksa2010a, Kohoutek2011, Rajapaksa2011a, Lu2011a, Lu2014}, we get $F_{\rm min} = 0.3$ pN and $F'_{\rm min} = 0.02$  pN/nm.
This makes collective particle excitation induced force detectable under typical laser intensities below the sample damage threshold. 
In contrast force detection due to molecular electronic and vibrational resonance is two orders of magnitude below the detection limit of conventional room temperature AFM.

The effect due to thermal expansion $\Delta h$ can be estimated based on $\Delta h = \alpha_T \Delta T $ \cite{Lu2014}.
Assuming a sample thickness of 10 nm, typical thermal expansion coefficients $\alpha_T = 10^{-4} - 10^{-5}$/K,  and a sample temperature increase of a few K after laser illumination \cite{Downes2006, Lu2014}, we get a thermal expansion value of $\Delta h =$ 0.1 -- 100 pm.
When this expansion of the sample is modulated at a resonance frequency of the cantilever, the corresponding force becomes $F_{\rm thermal} = k \Delta h \times Q$, 
where $k$ is the cantilever spring constant of typically 3 N/m, and $Q = $100 -- 200. The resulting force is in the range of $F_{\rm thermal} = 10$ pN -- 1 nN.
Thus the thermal expansion effect can readily be measured with AFM as shown experimentally even down to monolayer sensitivity with a single chemical bond expansion of few pm \cite{Lu2014}.

In previous attempts to detect the optical gradient force, a large force magnitude of $F \sim 2$ pN and an excellent agreement of the force spectrum with the far-field molecular absorption spectrum \cite{Rajapaksa2010a, Jahng2014} strongly suggest that the observed experimental results were in fact due to thermal expansion and not due to the optical gradient force as claimed.
Optical gradient force detection has also been assigned to the force contrast on bowtie antennas reaching few pN on resonance \cite{Kohoutek2011}. 
This assignment is feasible in principle according to our predictions. 
However, a force due to thermal expansion of similar magnitude is expected, leaving the underlying mechanism of the imaging contrast unclear.

In practice, the differences both in distance dependence and spectral frequency dependence of the optical gradient force and thermal expansion can be used to differentiate the two mechanisms.
Notably, the optical gradient force is a longer range effect, determined by the spatial extent of the near-field of the tip apex, and thus follows a continuous change with distance as determined by the tip radius (Fig.~\ref{PMMA_spec}f and Fig.~\ref{dis_dep}c).
In contrast, the force exerted on the tip through sample expansion requires a direct physical tip-sample contact, and should decay on even sub-nm distances above the sample, independent of tip radius. 

A way to differentiate the two mechanisms in imaging plasmonic optical antennas is to map the spatial force distribution.
The optical force is proportional to the local optical electric field, while the thermal expansion/absorption is due to resistive heating associated with electric currents. Thus distinct spatial maps result for the two different mechanisms since the current distribution peaks at the positions of minima in the electric field in optical antennas \cite{Olmon2010b}.

Fundamentally, the optical gradient force due to optical polarization and thermal expansion due to energy dissipation are two dynamic processes occurring on different time scales. 
This provides an additional opportunity for their distinction.
Optical polarization with coherent excitation is induced nearly instantaneously in the fs to ps range, as determined by the spectral line width. In contrast, thermalization of an optical excitation underlying optical absorpton leads to thermal expansion on ns time scales.

In summary, the dispersive line shape in probing electronic and vibrational materials resonances as related to the real part of the dielectric function of the sample can serve as a distinguishing attribute in optical gradient force microscopy. 
Albeit weak with forces in the fN to pN range, their detection with advanced atomic force techniques can provide for a novel form of optical scanning probe nano-imaging and -spectroscopy.
Compatible with a wide range of spectroscopies, including coherent and ultrafast techniques, it can complement related all-optical scattering scanning near-field microscopy (s-SNOM).
Our theory provides for the general description of the key parameters of strength, distance dependence, and spectral behavior for a simplified tip geometry, yet can readily be extended to specific tip geometries and other optical processes, including inelastic, {\it e.g.}, Raman, as well as nonlinear excitations.

\section{Acknowledgement}
Funding from the National Science Foundation (NSF Grant CHE 1306398) is gratefully acknowledged.
We thank Prof. M. A. Belkin and H. K. Wickramasinghe for valuable discussions.
%

%

\begin{thebibliography}{27}%
\makeatletter
\providecommand \@ifxundefined [1]{%
 \@ifx{#1\undefined}
}%
\providecommand \@ifnum [1]{%
 \ifnum #1\expandafter \@firstoftwo
 \else \expandafter \@secondoftwo
 \fi
}%
\providecommand \@ifx [1]{%
 \ifx #1\expandafter \@firstoftwo
 \else \expandafter \@secondoftwo
 \fi
}%
\providecommand \natexlab [1]{#1}%
\providecommand \enquote  [1]{``#1''}%
\providecommand \bibnamefont  [1]{#1}%
\providecommand \bibfnamefont [1]{#1}%
\providecommand \citenamefont [1]{#1}%
\providecommand \href@noop [0]{\@secondoftwo}%
\providecommand \href [0]{\begingroup \@sanitize@url \@href}%
\providecommand \@href[1]{\@@startlink{#1}\@@href}%
\providecommand \@@href[1]{\endgroup#1\@@endlink}%
\providecommand \@sanitize@url [0]{\catcode `\\12\catcode `\$12\catcode
  `\&12\catcode `\#12\catcode `\^12\catcode `\_12\catcode `\%12\relax}%
\providecommand \@@startlink[1]{}%
\providecommand \@@endlink[0]{}%
\providecommand \url  [0]{\begingroup\@sanitize@url \@url }%
\providecommand \@url [1]{\endgroup\@href {#1}{\urlprefix }}%
\providecommand \urlprefix  [0]{URL }%
\providecommand \Eprint [0]{\href }%
\providecommand \doibase [0]{http://dx.doi.org/}%
\providecommand \selectlanguage [0]{\@gobble}%
\providecommand \bibinfo  [0]{\@secondoftwo}%
\providecommand \bibfield  [0]{\@secondoftwo}%
\providecommand \translation [1]{[#1]}%
\providecommand \BibitemOpen [0]{}%
\providecommand \bibitemStop [0]{}%
\providecommand \bibitemNoStop [0]{.\EOS\space}%
\providecommand \EOS [0]{\spacefactor3000\relax}%
\providecommand \BibitemShut  [1]{\csname bibitem#1\endcsname}%
\let\auto@bib@innerbib\@empty

\bibitem [{\citenamefont {Anetsberger}\ \emph {et~al.}(2009)\citenamefont
  {Anetsberger}, \citenamefont {Arcizet}, \citenamefont {Gavartin},
  \citenamefont {Unterreithmeier}, \citenamefont {Weig}, \citenamefont
  {Kotthaus},\ and\ \citenamefont {Kippenberg}}]{Anetsberger2009}%
  \BibitemOpen
  \bibfield  {author} {\bibinfo {author} {\bibfnamefont {G.}~\bibnamefont
  {Anetsberger}}, \bibinfo {author} {\bibfnamefont {O.}~\bibnamefont
  {Arcizet}}, \bibinfo {author} {\bibfnamefont {E.}~\bibnamefont {Gavartin}},
  \bibinfo {author} {\bibfnamefont {Q.}~\bibnamefont {Unterreithmeier}},
  \bibinfo {author} {\bibfnamefont {E.}~\bibnamefont {Weig}}, \bibinfo {author}
  {\bibfnamefont {J.}~\bibnamefont {Kotthaus}}, \ and\ \bibinfo {author}
  {\bibfnamefont {T.}~\bibnamefont {Kippenberg}},\ }\href {\doibase
  10.1038/nphys1425} {\bibfield  {journal} {\bibinfo  {journal} {Nature
  Physics}\ }\textbf {\bibinfo {volume} {5}},\ \bibinfo {pages} {909} (\bibinfo
  {year} {2009})}\BibitemShut {NoStop}%
\bibitem [{\citenamefont {Berthelot}\ \emph {et~al.}(2014)\citenamefont
  {Berthelot}, \citenamefont {A\'{c}imovi\'{c}}, \citenamefont {Juan},
  \citenamefont {Kreuzer}, \citenamefont {Renger},\ and\ \citenamefont
  {Quidant}}]{Berthelot2014}%
  \BibitemOpen
  \bibfield  {author} {\bibinfo {author} {\bibfnamefont {J.}~\bibnamefont
  {Berthelot}}, \bibinfo {author} {\bibfnamefont {S.~S.}\ \bibnamefont
  {A\'{c}imovi\'{c}}}, \bibinfo {author} {\bibfnamefont {M.~L.}\ \bibnamefont
  {Juan}}, \bibinfo {author} {\bibfnamefont {M.~P.}\ \bibnamefont {Kreuzer}},
  \bibinfo {author} {\bibfnamefont {J.}~\bibnamefont {Renger}}, \ and\ \bibinfo
  {author} {\bibfnamefont {R.}~\bibnamefont {Quidant}},\ }\href {\doibase
  10.1038/nnano.2014.24} {\bibfield  {journal} {\bibinfo  {journal} {Nature
  nanotechnology}\ }\textbf {\bibinfo {volume} {9}},\ \bibinfo {pages} {295}
  (\bibinfo {year} {2014})}\BibitemShut {NoStop}%
\bibitem [{\citenamefont {Giessibl}(2003)}]{Giessibl2003}%
  \BibitemOpen
  \bibfield  {author} {\bibinfo {author} {\bibfnamefont {F.~J.}\ \bibnamefont
  {Giessibl}},\ }\href {\doibase 10.1103/RevModPhys.75.949} {\bibfield
  {journal} {\bibinfo  {journal} {Reviews of Modern Physics}\ }\textbf
  {\bibinfo {volume} {75}},\ \bibinfo {pages} {949} (\bibinfo {year}
  {2003})}\BibitemShut {NoStop}%
\bibitem [{\citenamefont {Emmrich}\ \emph {et~al.}(2015)\citenamefont
  {Emmrich}, \citenamefont {Huber}, \citenamefont {Pielmeier}, \citenamefont
  {Welker}, \citenamefont {Hofmann}, \citenamefont {Schneiderbauer},
  \citenamefont {Meuer}, \citenamefont {Polesya}, \citenamefont {Mankovsky},
  \citenamefont {Kodderitzsch}, \citenamefont {Ebert},\ and\ \citenamefont
  {Giessibl}}]{Emmrich2015}%
  \BibitemOpen
  \bibfield  {author} {\bibinfo {author} {\bibfnamefont {M.}~\bibnamefont
  {Emmrich}}, \bibinfo {author} {\bibfnamefont {F.}~\bibnamefont {Huber}},
  \bibinfo {author} {\bibfnamefont {F.}~\bibnamefont {Pielmeier}}, \bibinfo
  {author} {\bibfnamefont {J.}~\bibnamefont {Welker}}, \bibinfo {author}
  {\bibfnamefont {T.}~\bibnamefont {Hofmann}}, \bibinfo {author} {\bibfnamefont
  {M.}~\bibnamefont {Schneiderbauer}}, \bibinfo {author} {\bibfnamefont
  {D.}~\bibnamefont {Meuer}}, \bibinfo {author} {\bibfnamefont
  {S.}~\bibnamefont {Polesya}}, \bibinfo {author} {\bibfnamefont
  {S.}~\bibnamefont {Mankovsky}}, \bibinfo {author} {\bibfnamefont
  {D.}~\bibnamefont {Kodderitzsch}}, \bibinfo {author} {\bibfnamefont
  {H.}~\bibnamefont {Ebert}}, \ and\ \bibinfo {author} {\bibfnamefont {F.~J.}\
  \bibnamefont {Giessibl}},\ }\href {\doibase 10.1126/science.aaa5329}
  {\bibfield  {journal} {\bibinfo  {journal} {Science}\ }\textbf {\bibinfo
  {volume} {348}},\ \bibinfo {pages} {308} (\bibinfo {year}
  {2015})}\BibitemShut {NoStop}%
\bibitem [{\citenamefont {Betzig}\ \emph {et~al.}(1991)\citenamefont {Betzig},
  \citenamefont {Trautman}, \citenamefont {Harris}, \citenamefont {Weiner},\
  and\ \citenamefont {Kostelak}}]{Betzig1991}%
  \BibitemOpen
  \bibfield  {author} {\bibinfo {author} {\bibfnamefont {E.}~\bibnamefont
  {Betzig}}, \bibinfo {author} {\bibfnamefont {J.~K.}\ \bibnamefont
  {Trautman}}, \bibinfo {author} {\bibfnamefont {T.~D.}\ \bibnamefont
  {Harris}}, \bibinfo {author} {\bibfnamefont {J.~S.}\ \bibnamefont {Weiner}},
  \ and\ \bibinfo {author} {\bibfnamefont {R.~L.}\ \bibnamefont {Kostelak}},\
  }\href {\doibase 10.1126/science.251.5000.1468} {\bibfield  {journal}
  {\bibinfo  {journal} {Science}\ }\textbf {\bibinfo {volume} {251}},\ \bibinfo
  {pages} {1468} (\bibinfo {year} {1991})}\BibitemShut {NoStop}%
\bibitem [{\citenamefont {Zenhausern}\ \emph {et~al.}(1995)\citenamefont
  {Zenhausern}, \citenamefont {Martin},\ and\ \citenamefont
  {Wickramasinghe}}]{Zenhausern1995}%
  \BibitemOpen
  \bibfield  {author} {\bibinfo {author} {\bibfnamefont {F.}~\bibnamefont
  {Zenhausern}}, \bibinfo {author} {\bibfnamefont {Y.}~\bibnamefont {Martin}},
  \ and\ \bibinfo {author} {\bibfnamefont {H.~K.}\ \bibnamefont
  {Wickramasinghe}},\ }\href {\doibase 10.1126/science.269.5227.1083}
  {\bibfield  {journal} {\bibinfo  {journal} {Science}\ }\textbf {\bibinfo
  {volume} {269}},\ \bibinfo {pages} {1083} (\bibinfo {year}
  {1995})}\BibitemShut {NoStop}%
\bibitem [{\citenamefont {Depasse}\ and\ \citenamefont
  {Courjon}(1992)}]{Depasse1992}%
  \BibitemOpen
  \bibfield  {author} {\bibinfo {author} {\bibfnamefont {F.}~\bibnamefont
  {Depasse}}\ and\ \bibinfo {author} {\bibfnamefont {D.}~\bibnamefont
  {Courjon}},\ }\href {\doibase 10.1016/0030-4018(92)90383-3} {\bibfield
  {journal} {\bibinfo  {journal} {Optics Communications}\ }\textbf {\bibinfo
  {volume} {87}},\ \bibinfo {pages} {79} (\bibinfo {year} {1992})}\BibitemShut
  {NoStop}%
\bibitem [{\citenamefont {Zhu}\ \emph {et~al.}(1997)\citenamefont {Zhu},
  \citenamefont {Huang}, \citenamefont {Zhou}, \citenamefont {Yang},
  \citenamefont {Wang}, \citenamefont {Ling}, \citenamefont {Dai},\ and\
  \citenamefont {Gan}}]{Zhu1997}%
  \BibitemOpen
  \bibfield  {author} {\bibinfo {author} {\bibfnamefont {X.}~\bibnamefont
  {Zhu}}, \bibinfo {author} {\bibfnamefont {G.-S.}\ \bibnamefont {Huang}},
  \bibinfo {author} {\bibfnamefont {H.-T.}\ \bibnamefont {Zhou}}, \bibinfo
  {author} {\bibfnamefont {X.}~\bibnamefont {Yang}}, \bibinfo {author}
  {\bibfnamefont {Z.}~\bibnamefont {Wang}}, \bibinfo {author} {\bibfnamefont
  {Y.}~\bibnamefont {Ling}}, \bibinfo {author} {\bibfnamefont {Y.-D.}\
  \bibnamefont {Dai}}, \ and\ \bibinfo {author} {\bibfnamefont {Z.-Z.}\
  \bibnamefont {Gan}},\ }\href {\doibase 10.1007/BF02936034} {\bibfield
  {journal} {\bibinfo  {journal} {Optical Review}\ }\textbf {\bibinfo {volume}
  {4}},\ \bibinfo {pages} {A236} (\bibinfo {year} {1997})}\BibitemShut
  {NoStop}%
\bibitem [{\citenamefont {Kohlgraf-Owens}\ \emph {et~al.}(2014)\citenamefont
  {Kohlgraf-Owens}, \citenamefont {Greusard}, \citenamefont {Sukhov},
  \citenamefont {Wilde},\ and\ \citenamefont {Dogariu}}]{Kohlgraf-Owens2014}%
  \BibitemOpen
  \bibfield  {author} {\bibinfo {author} {\bibfnamefont {D.~C.}\ \bibnamefont
  {Kohlgraf-Owens}}, \bibinfo {author} {\bibfnamefont {L.}~\bibnamefont
  {Greusard}}, \bibinfo {author} {\bibfnamefont {S.}~\bibnamefont {Sukhov}},
  \bibinfo {author} {\bibfnamefont {Y.~D.}\ \bibnamefont {Wilde}}, \ and\
  \bibinfo {author} {\bibfnamefont {A.}~\bibnamefont {Dogariu}},\ }\href
  {\doibase 10.1088/0957-4484/25/3/035203} {\bibfield  {journal} {\bibinfo
  {journal} {Nanotechnology}\ }\textbf {\bibinfo {volume} {25}},\ \bibinfo
  {pages} {035203} (\bibinfo {year} {2014})}\BibitemShut {NoStop}%
\bibitem [{\citenamefont {Satoh}\ \emph {et~al.}(2010)\citenamefont {Satoh},
  \citenamefont {Fukuma}, \citenamefont {Kobayashi}, \citenamefont {Watanabe},
  \citenamefont {Fujii}, \citenamefont {Matsushige},\ and\ \citenamefont
  {Yamada}}]{Satoh2010}%
  \BibitemOpen
  \bibfield  {author} {\bibinfo {author} {\bibfnamefont {N.}~\bibnamefont
  {Satoh}}, \bibinfo {author} {\bibfnamefont {T.}~\bibnamefont {Fukuma}},
  \bibinfo {author} {\bibfnamefont {K.}~\bibnamefont {Kobayashi}}, \bibinfo
  {author} {\bibfnamefont {S.}~\bibnamefont {Watanabe}}, \bibinfo {author}
  {\bibfnamefont {T.}~\bibnamefont {Fujii}}, \bibinfo {author} {\bibfnamefont
  {K.}~\bibnamefont {Matsushige}}, \ and\ \bibinfo {author} {\bibfnamefont
  {H.}~\bibnamefont {Yamada}},\ }\href {\doibase 10.1063/1.3449131} {\bibfield
  {journal} {\bibinfo  {journal} {Applied Physics Letters}\ }\textbf {\bibinfo
  {volume} {96}},\ \bibinfo {pages} {3} (\bibinfo {year} {2010})}\BibitemShut
  {NoStop}%
\bibitem [{\citenamefont {Rajapaksa}\ \emph {et~al.}(2010)\citenamefont
  {Rajapaksa}, \citenamefont {Uenal},\ and\ \citenamefont
  {Wickramasinghe}}]{Rajapaksa2010a}%
  \BibitemOpen
  \bibfield  {author} {\bibinfo {author} {\bibfnamefont {I.}~\bibnamefont
  {Rajapaksa}}, \bibinfo {author} {\bibfnamefont {K.}~\bibnamefont {Uenal}}, \
  and\ \bibinfo {author} {\bibfnamefont {H.~K.}\ \bibnamefont
  {Wickramasinghe}},\ }\href {\doibase 10.1063/1.3480608} {\bibfield  {journal}
  {\bibinfo  {journal} {Applied Physics Letters}\ }\textbf {\bibinfo {volume}
  {97}},\ \bibinfo {pages} {2010} (\bibinfo {year} {2010})}\BibitemShut
  {NoStop}%
\bibitem [{\citenamefont {Rajapaksa}\ and\ \citenamefont {{Kumar
  Wickramasinghe}}(2011)}]{Rajapaksa2011a}%
  \BibitemOpen
  \bibfield  {author} {\bibinfo {author} {\bibfnamefont {I.}~\bibnamefont
  {Rajapaksa}}\ and\ \bibinfo {author} {\bibfnamefont {H.}~\bibnamefont {{Kumar
  Wickramasinghe}}},\ }\href {\doibase 10.1063/1.3652760} {\bibfield  {journal}
  {\bibinfo  {journal} {Applied Physics Letters}\ }\textbf {\bibinfo {volume}
  {99}},\ \bibinfo {pages} {161103} (\bibinfo {year} {2011})}\BibitemShut
  {NoStop}%
\bibitem [{\citenamefont {{De Angelis}}\ \emph {et~al.}(2012)\citenamefont {{De
  Angelis}}, \citenamefont {Zaccaria},\ and\ \citenamefont {{Di
  Fabrizio}}}]{DeAngelis2012}%
  \BibitemOpen
  \bibfield  {author} {\bibinfo {author} {\bibfnamefont {F.}~\bibnamefont {{De
  Angelis}}}, \bibinfo {author} {\bibfnamefont {R.~P.}\ \bibnamefont
  {Zaccaria}}, \ and\ \bibinfo {author} {\bibfnamefont {E.}~\bibnamefont {{Di
  Fabrizio}}},\ }\href {\doibase 10.1364/OE.20.029626} {\bibfield  {journal}
  {\bibinfo  {journal} {Optics Express}\ }\textbf {\bibinfo {volume} {20}},\
  \bibinfo {pages} {29626} (\bibinfo {year} {2012})}\BibitemShut {NoStop}%
\bibitem [{\citenamefont {Jahng}\ \emph {et~al.}(2014)\citenamefont {Jahng},
  \citenamefont {Brocious}, \citenamefont {Fishman}, \citenamefont {Huang},
  \citenamefont {Li}, \citenamefont {Tamma}, \citenamefont {Wickramasinghe},\
  and\ \citenamefont {Potma}}]{Jahng2014}%
  \BibitemOpen
  \bibfield  {author} {\bibinfo {author} {\bibfnamefont {J.}~\bibnamefont
  {Jahng}}, \bibinfo {author} {\bibfnamefont {J.}~\bibnamefont {Brocious}},
  \bibinfo {author} {\bibfnamefont {D.~A.}\ \bibnamefont {Fishman}}, \bibinfo
  {author} {\bibfnamefont {F.}~\bibnamefont {Huang}}, \bibinfo {author}
  {\bibfnamefont {X.}~\bibnamefont {Li}}, \bibinfo {author} {\bibfnamefont
  {V.~A.}\ \bibnamefont {Tamma}}, \bibinfo {author} {\bibfnamefont {H.~K.}\
  \bibnamefont {Wickramasinghe}}, \ and\ \bibinfo {author} {\bibfnamefont
  {E.~O.}\ \bibnamefont {Potma}},\ }\href {\doibase 10.1103/PhysRevB.90.155417}
  {\bibfield  {journal} {\bibinfo  {journal} {Physical Review B}\ }\textbf
  {\bibinfo {volume} {90}},\ \bibinfo {pages} {155417} (\bibinfo {year}
  {2014})}\BibitemShut {NoStop}%
\bibitem [{\citenamefont {Dazzi}\ \emph {et~al.}(2005)\citenamefont {Dazzi},
  \citenamefont {Prazeres}, \citenamefont {Glotin},\ and\ \citenamefont
  {Ortega}}]{Dazzi2005}%
  \BibitemOpen
  \bibfield  {author} {\bibinfo {author} {\bibfnamefont {A.}~\bibnamefont
  {Dazzi}}, \bibinfo {author} {\bibfnamefont {R.}~\bibnamefont {Prazeres}},
  \bibinfo {author} {\bibfnamefont {F.}~\bibnamefont {Glotin}}, \ and\ \bibinfo
  {author} {\bibfnamefont {J.~M.}\ \bibnamefont {Ortega}},\ }\href {\doibase
  10.1364/OL.30.002388} {\bibfield  {journal} {\bibinfo  {journal} {Optics
  Letters}\ }\textbf {\bibinfo {volume} {30}},\ \bibinfo {pages} {2388}
  (\bibinfo {year} {2005})}\BibitemShut {NoStop}%
\bibitem [{\citenamefont {Dazzi}\ \emph {et~al.}(2012)\citenamefont {Dazzi},
  \citenamefont {Prater}, \citenamefont {Hu}, \citenamefont {Chase},
  \citenamefont {Rabolt},\ and\ \citenamefont {Marcott}}]{Dazzi2012}%
  \BibitemOpen
  \bibfield  {author} {\bibinfo {author} {\bibfnamefont {A.}~\bibnamefont
  {Dazzi}}, \bibinfo {author} {\bibfnamefont {C.~B.}\ \bibnamefont {Prater}},
  \bibinfo {author} {\bibfnamefont {Q.}~\bibnamefont {Hu}}, \bibinfo {author}
  {\bibfnamefont {D.~B.}\ \bibnamefont {Chase}}, \bibinfo {author}
  {\bibfnamefont {J.~F.}\ \bibnamefont {Rabolt}}, \ and\ \bibinfo {author}
  {\bibfnamefont {C.}~\bibnamefont {Marcott}},\ }\href {\doibase
  10.1366/12-06804} {\bibfield  {journal} {\bibinfo  {journal} {Applied
  Spectroscopy}\ }\textbf {\bibinfo {volume} {66 N62}},\ \bibinfo {pages}
  {1365} (\bibinfo {year} {2012})}\BibitemShut {NoStop}%
\bibitem [{\citenamefont {Lu}\ \emph {et~al.}(2014)\citenamefont {Lu},
  \citenamefont {Jin},\ and\ \citenamefont {Belkin}}]{Lu2014}%
  \BibitemOpen
  \bibfield  {author} {\bibinfo {author} {\bibfnamefont {F.}~\bibnamefont
  {Lu}}, \bibinfo {author} {\bibfnamefont {M.}~\bibnamefont {Jin}}, \ and\
  \bibinfo {author} {\bibfnamefont {M.~A.}\ \bibnamefont {Belkin}},\ }\href
  {\doibase 10.1038/nphoton.2013.373} {\bibfield  {journal} {\bibinfo
  {journal} {Nature Photonics}\ }\textbf {\bibinfo {volume} {8}},\ \bibinfo
  {pages} {307} (\bibinfo {year} {2014})}\BibitemShut {NoStop}%
\bibitem [{\citenamefont {Kohoutek}\ \emph {et~al.}(2011)\citenamefont
  {Kohoutek}, \citenamefont {Dey}, \citenamefont {Bonakdar}, \citenamefont
  {Gelfand}, \citenamefont {Sklar}, \citenamefont {Memis},\ and\ \citenamefont
  {Mohseni}}]{Kohoutek2011}%
  \BibitemOpen
  \bibfield  {author} {\bibinfo {author} {\bibfnamefont {J.}~\bibnamefont
  {Kohoutek}}, \bibinfo {author} {\bibfnamefont {D.}~\bibnamefont {Dey}},
  \bibinfo {author} {\bibfnamefont {A.}~\bibnamefont {Bonakdar}}, \bibinfo
  {author} {\bibfnamefont {R.}~\bibnamefont {Gelfand}}, \bibinfo {author}
  {\bibfnamefont {A.}~\bibnamefont {Sklar}}, \bibinfo {author} {\bibfnamefont
  {O.~G.}\ \bibnamefont {Memis}}, \ and\ \bibinfo {author} {\bibfnamefont
  {H.}~\bibnamefont {Mohseni}},\ }\href {\doibase 10.1021/nl201780y} {\bibfield
   {journal} {\bibinfo  {journal} {Nano Letters}\ }\textbf {\bibinfo {volume}
  {11}},\ \bibinfo {pages} {3378} (\bibinfo {year} {2011})}\BibitemShut
  {NoStop}%
\bibitem [{\citenamefont {Katzenmeyer}\ \emph {et~al.}(2014)\citenamefont
  {Katzenmeyer}, \citenamefont {Chae}, \citenamefont {Kasica}, \citenamefont
  {Holland}, \citenamefont {Lahiri},\ and\ \citenamefont
  {Centrone}}]{Katzenmeyer2014}%
  \BibitemOpen
  \bibfield  {author} {\bibinfo {author} {\bibfnamefont {A.~M.}\ \bibnamefont
  {Katzenmeyer}}, \bibinfo {author} {\bibfnamefont {J.}~\bibnamefont {Chae}},
  \bibinfo {author} {\bibfnamefont {R.}~\bibnamefont {Kasica}}, \bibinfo
  {author} {\bibfnamefont {G.}~\bibnamefont {Holland}}, \bibinfo {author}
  {\bibfnamefont {B.}~\bibnamefont {Lahiri}}, \ and\ \bibinfo {author}
  {\bibfnamefont {A.}~\bibnamefont {Centrone}},\ }\href {\doibase
  10.1002/adom.201400005} {\bibfield  {journal} {\bibinfo  {journal} {Advanced
  Optical Materials}\ }\textbf {\bibinfo {volume} {2}},\ \bibinfo {pages} {718}
  (\bibinfo {year} {2014})}\BibitemShut {NoStop}%
\bibitem [{\citenamefont {Jahng}\ \emph {et~al.}(2015)\citenamefont {Jahng},
  \citenamefont {Brocious}, \citenamefont {Fishman}, \citenamefont {Yampolsky},
  \citenamefont {Nowak}, \citenamefont {Huang}, \citenamefont {Apkarian},
  \citenamefont {Wickramasinghe},\ and\ \citenamefont {Potma}}]{Jahng2015}%
  \BibitemOpen
  \bibfield  {author} {\bibinfo {author} {\bibfnamefont {J.}~\bibnamefont
  {Jahng}}, \bibinfo {author} {\bibfnamefont {J.}~\bibnamefont {Brocious}},
  \bibinfo {author} {\bibfnamefont {D.~A.}\ \bibnamefont {Fishman}}, \bibinfo
  {author} {\bibfnamefont {S.}~\bibnamefont {Yampolsky}}, \bibinfo {author}
  {\bibfnamefont {D.}~\bibnamefont {Nowak}}, \bibinfo {author} {\bibfnamefont
  {F.}~\bibnamefont {Huang}}, \bibinfo {author} {\bibfnamefont {V.~A.}\
  \bibnamefont {Apkarian}}, \bibinfo {author} {\bibfnamefont {H.~K.}\
  \bibnamefont {Wickramasinghe}}, \ and\ \bibinfo {author} {\bibfnamefont
  {E.~O.}\ \bibnamefont {Potma}},\ }\href {\doibase 10.1063/1.4913853}
  {\bibfield  {journal} {\bibinfo  {journal} {Applied Physics Letters}\
  }\textbf {\bibinfo {volume} {106}},\ \bibinfo {pages} {083113} (\bibinfo
  {year} {2015})}\BibitemShut {NoStop}%
\bibitem [{\citenamefont {Huang}\ \emph {et~al.}(2015)\citenamefont {Huang},
  \citenamefont {{Ananth Tamma}}, \citenamefont {Mardy}, \citenamefont
  {Burdett},\ and\ \citenamefont {Wickramasinghe}}]{Huang2015}%
  \BibitemOpen
  \bibfield  {author} {\bibinfo {author} {\bibfnamefont {F.}~\bibnamefont
  {Huang}}, \bibinfo {author} {\bibfnamefont {V.}~\bibnamefont {{Ananth
  Tamma}}}, \bibinfo {author} {\bibfnamefont {Z.}~\bibnamefont {Mardy}},
  \bibinfo {author} {\bibfnamefont {J.}~\bibnamefont {Burdett}}, \ and\
  \bibinfo {author} {\bibfnamefont {H.~K.}\ \bibnamefont {Wickramasinghe}},\
  }\href {\doibase 10.1038/srep10610} {\bibfield  {journal} {\bibinfo
  {journal} {Scientific Reports}\ }\textbf {\bibinfo {volume} {5}},\ \bibinfo
  {pages} {10610} (\bibinfo {year} {2015})}\BibitemShut {NoStop}%
\bibitem [{\citenamefont {Berweger}\ \emph {et~al.}(2013)\citenamefont
  {Berweger}, \citenamefont {Nguyen}, \citenamefont {Muller}, \citenamefont
  {Bechtel}, \citenamefont {Perkins},\ and\ \citenamefont
  {Raschke}}]{Berweger2013b}%
  \BibitemOpen
  \bibfield  {author} {\bibinfo {author} {\bibfnamefont {S.}~\bibnamefont
  {Berweger}}, \bibinfo {author} {\bibfnamefont {D.~M.}\ \bibnamefont
  {Nguyen}}, \bibinfo {author} {\bibfnamefont {E.~A.}\ \bibnamefont {Muller}},
  \bibinfo {author} {\bibfnamefont {H.~A.}\ \bibnamefont {Bechtel}}, \bibinfo
  {author} {\bibfnamefont {T.~T.}\ \bibnamefont {Perkins}}, \ and\ \bibinfo
  {author} {\bibfnamefont {M.~B.}\ \bibnamefont {Raschke}},\ }\href {\doibase
  10.1021/ja409815g} {\bibfield  {journal} {\bibinfo  {journal} {Journal of the
  American Chemical Society}\ }\textbf {\bibinfo {volume} {135}},\ \bibinfo
  {pages} {18292} (\bibinfo {year} {2013})}\BibitemShut {NoStop}%
\bibitem [{\citenamefont {Romero}\ \emph {et~al.}(2006)\citenamefont {Romero},
  \citenamefont {Aizpurua}, \citenamefont {Bryant},\ and\ \citenamefont
  {{Garc\'{\i}a De Abajo}}}]{Romero2006}%
  \BibitemOpen
  \bibfield  {author} {\bibinfo {author} {\bibfnamefont {I.}~\bibnamefont
  {Romero}}, \bibinfo {author} {\bibfnamefont {J.}~\bibnamefont {Aizpurua}},
  \bibinfo {author} {\bibfnamefont {G.~W.}\ \bibnamefont {Bryant}}, \ and\
  \bibinfo {author} {\bibfnamefont {F.~J.}\ \bibnamefont {{Garc\'{\i}a De
  Abajo}}},\ }\href {\doibase 10.1364/OE.14.009988} {\bibfield  {journal}
  {\bibinfo  {journal} {Optics Express}\ }\textbf {\bibinfo {volume} {14}},\
  \bibinfo {pages} {9988} (\bibinfo {year} {2006})}\BibitemShut {NoStop}%
\bibitem [{\citenamefont {Martin}\ \emph {et~al.}(1987)\citenamefont {Martin},
  \citenamefont {Williams},\ and\ \citenamefont {Wickramasinghe}}]{Martin1987}%
  \BibitemOpen
  \bibfield  {author} {\bibinfo {author} {\bibfnamefont {Y.}~\bibnamefont
  {Martin}}, \bibinfo {author} {\bibfnamefont {C.~C.}\ \bibnamefont
  {Williams}}, \ and\ \bibinfo {author} {\bibfnamefont {H.~K.}\ \bibnamefont
  {Wickramasinghe}},\ }\href {\doibase 10.1063/1.338807} {\bibfield  {journal}
  {\bibinfo  {journal} {Journal of Applied Physics}\ }\textbf {\bibinfo
  {volume} {61}},\ \bibinfo {pages} {4723} (\bibinfo {year}
  {1987})}\BibitemShut {NoStop}%
\bibitem [{\citenamefont {Lu}\ and\ \citenamefont {Belkin}(2011)}]{Lu2011a}%
  \BibitemOpen
  \bibfield  {author} {\bibinfo {author} {\bibfnamefont {F.}~\bibnamefont
  {Lu}}\ and\ \bibinfo {author} {\bibfnamefont {M.~A.}\ \bibnamefont
  {Belkin}},\ }\href {http://www.ncbi.nlm.nih.gov/pubmed/21997003} {\bibfield
  {journal} {\bibinfo  {journal} {Optics Express}\ }\textbf {\bibinfo {volume}
  {19}},\ \bibinfo {pages} {19942} (\bibinfo {year} {2011})}\BibitemShut
  {NoStop}%
\bibitem [{\citenamefont {Downes}\ \emph {et~al.}(2006)\citenamefont {Downes},
  \citenamefont {Salter},\ and\ \citenamefont {Elfick}}]{Downes2006}%
  \BibitemOpen
  \bibfield  {author} {\bibinfo {author} {\bibfnamefont {A.}~\bibnamefont
  {Downes}}, \bibinfo {author} {\bibfnamefont {D.}~\bibnamefont {Salter}}, \
  and\ \bibinfo {author} {\bibfnamefont {A.}~\bibnamefont {Elfick}},\ }\href
  {\doibase 10.1364/OE.14.005216} {\bibfield  {journal} {\bibinfo  {journal}
  {Optics express}\ }\textbf {\bibinfo {volume} {14}},\ \bibinfo {pages} {5216}
  (\bibinfo {year} {2006})}\BibitemShut {NoStop}%
\bibitem [{\citenamefont {Olmon}\ \emph {et~al.}(2010)\citenamefont {Olmon},
  \citenamefont {Rang}, \citenamefont {Krenz}, \citenamefont {Lail},
  \citenamefont {Saraf}, \citenamefont {Boreman},\ and\ \citenamefont
  {Raschke}}]{Olmon2010b}%
  \BibitemOpen
  \bibfield  {author} {\bibinfo {author} {\bibfnamefont {R.~L.}\ \bibnamefont
  {Olmon}}, \bibinfo {author} {\bibfnamefont {M.}~\bibnamefont {Rang}},
  \bibinfo {author} {\bibfnamefont {P.~M.}\ \bibnamefont {Krenz}}, \bibinfo
  {author} {\bibfnamefont {B.~A.}\ \bibnamefont {Lail}}, \bibinfo {author}
  {\bibfnamefont {L.~V.}\ \bibnamefont {Saraf}}, \bibinfo {author}
  {\bibfnamefont {G.~D.}\ \bibnamefont {Boreman}}, \ and\ \bibinfo {author}
  {\bibfnamefont {M.~B.}\ \bibnamefont {Raschke}},\ }\href {\doibase
  10.1103/PhysRevLett.105.167403} {\bibfield  {journal} {\bibinfo  {journal}
  {Physical Review Letters}\ }\textbf {\bibinfo {volume} {105}},\ \bibinfo
  {pages} {167403} (\bibinfo {year} {2010})}\BibitemShut {NoStop}%
\end{thebibliography}
\end{document}